\documentclass[aps,prl,numerical,linenumbers,superscriptaddress,showpacs,floatfix,reprint]{revtex4-1}
\usepackage[dvipdfm]{graphicx}
\usepackage{amsmath,amssymb,amsfonts}
\usepackage{color}

\newcommand{\be}{\begin{eqnarray}}
\newcommand{\ee}{\end{eqnarray}}

\def\v2{\mbox{$v_2$}}

\begin{document}


%
\title{Scaling of the higher-order flow harmonics: implications for \\
initial-eccentricity models and the ``viscous horizon''
}
%
%
%
\author{ Roy~A.~Lacey}
\email[E-mail: ]{Roy.Lacey@Stonybrook.edu}
\affiliation{Department of Chemistry, 
Stony Brook University, \\
Stony Brook, NY, 11794-3400, USA}
\author{A.~Taranenko}
\affiliation{Department of Chemistry, 
Stony Brook University, \\
Stony Brook, NY, 11794-3400, USA} 
\author{ J.~Jia}$^2$
\affiliation{Department of Chemistry, 
Stony Brook University, \\
Stony Brook, NY, 11794-3400, USA}
\affiliation{Physics Department, Bookhaven National Laboratory, \\
Upton, New York 11973-5000, USA}
\author{ N.~N.~Ajitanand} 
\affiliation{Department of Chemistry, 
Stony Brook University, \\
Stony Brook, NY, 11794-3400, USA}
\author{ J.~M.~Alexander}
\affiliation{Department of Chemistry, 
Stony Brook University, \\
Stony Brook, NY, 11794-3400, USA}



\date{\today}


\begin{abstract}

	The scaling properties of the flow harmonics for charged hadrons $v_{n}$ and their 
ratios $[ v_n/(v_2)^{n/2}]_{n\geq 3}$, are studied for a broad range of transverse 
momenta ($p_T$) and centrality selections in Au+Au 
and Pb+Pb collisions at $\sqrt{s_{NN}}=0.2 \text{ and } 2.76$ TeV respectively.
At relatively low $p_T$, these scaling properties are found to be compatible with the 
expected growth of viscous damping for sound propagation in the plasma produced in these 
collisions. They also provide important constraints for distinguishing between the two 
leading models of collision eccentricities, as well as a route to constrain the 
relaxation time and make estimates for the ratio of 
viscosity to entropy density $\eta/s$, and the ``viscous horizon'' or length-scale 
which characterizes the highest harmonic which survives viscous damping.

\end{abstract}


\maketitle


	Full characterization of the transport properties of the strongly interacting 
matter produced in heavy ion collisions, is a central goal of the 
experimental heavy ion programs at both the Relativistic Heavy Ion Collider (RHIC) and the 
Large Hadron Collider (LHC). Collective flow, as manifested 
by the anisotropic emission of particles in the plane transverse to the beam 
direction \cite{Lacey:2001va,*Snellings:2001nf},  
continues to play an important role in these ongoing efforts \cite{Gyulassy:2004zy,Molnar:2004yh,
Lacey:2006bc,Adare:2006nq,Romatschke:2007mq,
Xu:2007jv,Drescher:2007cd,Shuryak:2008eq,Luzum:2008cw,Song:2008hj,
Dusling:2007gi,Bozek:2009mz,Denicol:2010tr,Lacey:2010fe}. 
This anisotropy can be characterized, as a function of 
particle transverse momentum $p_T$ and collision centrality (cent) or the number 
of participant nucleons $N_{\text{part}}$, by the Fourier coefficients $v_n$;
\begin{equation}
\frac{dN}{d\phi} \propto \left( 1 + \sum_{n=1}2v_n\cos(n\phi - n\Psi_n) \right),
\label{eq:1}
\end{equation}  
and by the pair-wise distribution in the azimuthal angle difference 
($\Delta\phi =\phi_a - \phi_b$) between particle pairs with transverse 
momenta $p^a_{T}$ and $p^b_{T}$ (respectively) 
\cite{Lacey:2001va,Mocsy:2010um};
%
\begin{equation}
\frac{dN^{\text{pairs}}}{d\Delta\phi} \propto \left( 1 + \sum_{n=1}2v^a_nv^b_n\cos(n\Delta\phi) \right),
\label{eq:2}
\end{equation}
where $\phi$ is the azimuthal angle of an emitted particle, 
and $\Psi_{n}$ are the azimuths of the estimated participant event 
planes \cite{Ollitrault:1992bk,Adare:2010ux};
%
\begin{eqnarray}
v_n = \left\langle \cos n(\phi - \Psi_n) \right\rangle \,\,\, \nonumber \\
v^*_n = \left\langle \cos n(\phi - \Psi_m) \right\rangle, \, \, n \ne m,
\label{eq:3}
\end{eqnarray}
where the brackets denote averaging over particles and events.
Here, the the starred notation is used to distinguish the $n$-th order moments 
obtained relative to the $m$-th order event plane $\Psi_m$ (eg. $v^*_4$ relative to $\Psi_2$). 
For flow driven anisotropy devoid of non flow effects, the Fourier coefficients 
obtained with Eqs. \ref{eq:1} and \ref{eq:2} are equivalent.

Flow coefficients stem from an eccentricity-driven hydrodynamic expansion
of the matter in the collision zone \cite{Heinz:2001xi,Teaney:2003kp,Huovinen:2001cy, 
Hirano:2002ds,Romatschke:2007mq,Song:2008hj,Schenke:2010rr},
{\em i.e.}, a finite eccentricity $\varepsilon_n$ drives uneven pressure gradients in- and out 
of the event plane, and the resulting expansion leads to the anisotropic emission of particles 
about this plane.  The coefficients  $v_{n}(p_T,\text{cent})$ (for odd and even $n$) 
are sensitive to both the initial eccentricity and the specific shear viscosity 
$\eta/s$ ({\em i.e.}\ the ratio of shear viscosity $\eta$ to 
entropy density $s$) of the expanding hot plasma \cite{Heinz:2002rs,Teaney:2003kp,Lacey:2006pn,
Romatschke:2007mq,Drescher:2007cd,Xu:2007jv,Greco:2008fs}.
Here, it is noteworthy that, for symmetric systems, the 
symmetry transformation $\Psi_{RP} \rightarrow \Psi_{RP} + \pi$, dictates that 
the odd harmonics are zero for smooth ideal eccentricity profiles. However, the ``lumpy" 
transverse density distributions generated in individual collisions, can result in eccentricity 
profiles which have no particular symmetry, so the odd harmonics are 
not required to be zero from event to event. Fortuitously, the pervasive assumption 
of a smooth eccentricity profile has hindered full exploitation of the odd harmonics until
recently \cite{Alver:2010gr}.

	Because of the acoustic nature of anisotropic flow ({\em i.e.} it is driven by pressure gradients),
a transparent way to evaluate the strength of dissipative effects is to consider the 
attenuation of sound waves. In the presence of viscosity, sound intensity is 
exponentially damped $e^{(-r/\Gamma_s)}$ relative to the sound 
attenuation length $\Gamma_s$. This can be expressed in terms of a perturbation to 
the energy-momentum tensor $T_{\mu\nu}$ \cite{Staig:2010pn}:
\be
\delta T_{\mu\nu} (t) = \exp{\left(-\frac{2}{3} \frac{\eta}{s} k^2 \frac{t}{T } \right)}  \delta T_{\mu\nu} (0),
\label{eq:5}
\ee 
where viscous damping reflects the dispersion relation for sound propagation,
and the spectrum of initial (t = 0) perturbations can be associated with 
the harmonics of the shape deformations and density fluctuations. 
Here, $k$ is the wave number for these harmonics, and $t$ and $T$ are the 
expansion time and the temperature of the plasma respectively. 
For a collision zone of transverse size ${\bar R}$, Eq.~\ref{eq:5} indicates that 
viscous corrections for the eccentricity-driven flow harmonics with wavelengths 
$2\pi {\bar R}/n$ for $n\ge 1$ ({\em i.e.} $k \sim n/{\bar R}$), dampen 
exponentially as $n^2$.
The {\em ``viscous horizon''} or length scale $r_v = 2\pi{\bar{R}}/n_v$ is also 
linked to the order of the highest harmonic $n_v$ which effectively survives viscous damping. 
That is, it separates the high frequency sound modes which are fully damped from those which are 
not \cite{Staig:2010pn}.   
The sound horizon $r_s = \int^{\tau_f}_{\tau_0} d\tau c_s(\tau)$, or the distance 
sound travels at speed $c_s(\tau)$ until flow freeze-out $\tau_f$, sets the length scale 
for suppression of low frequency superhorizon modes with wavelengths 
$2\pi R_{\!f}/n > 2r_s$, where $R_{\!f}$ is the transverse size at sound freeze-out.   
Thus, the relative magnitudes of the flow harmonics $v_n$ can  
provide important constraints for pinning down the magnitude of the transport 
coefficients $\eta/s$ and $c_s$, as well as the ``correct'' model for eccentricity 
determinations \cite{Staig:2010pn,Lacey:2010hw,Alver:2010dn}.

	Viscous damping for sound propagation in the plasma does not indicate an explicit 
$p_T$ dependence for the relative magnitudes of $v_n$ (cf. Eq.~\ref{eq:5}). However, for 
a finite viscosity in the plasma, the resulting asymmetry in the energy-momentum tensor manifests 
as a correction to the local particle distribution ($f$) after freeze-out \cite{Dusling:2009df};
\begin{equation}
    f = f_0 + \delta f(p_T), 
\label{eq:6}
\end{equation}
where $f_0$ is the equilibrium distribution and $\delta f(p_T)$ is its first order correction. 
The latter acts as a viscous correction and is known to reduce the magnitude of  
$v_2(p_T)$, especially for $p_T \agt 0.7$ GeV/c \cite{Dusling:2009df}.
The relative magnitudes of $v_n(p_T)$ are expected to be dominated by the dispersion 
relation for sound propagation, albeit with some influence 
from $\delta f(p_T)$. For relatively small values of $\eta/s$, this influence 
on the $p_T$-dependent viscous corrections would also be small. 
Thus, a characteristic scaling relationship between $v_{n, n\geq 3}(p_T)$ 
and $v_2(p_T)$ might be expected.

In this letter, we investigate the scaling properties of $v_n(p_T,\text{cent})$ and the 
ratios $\left[v_n(p_T)/(v_2(p_T))^{n/2}\right]_{n \geq 3}$ for charged hadrons produced 
in ultrarelativistic Au+Au and Pb+Pb collisions. We find scaling patterns that: 
(i) validate the viscous damping expected for sound propagation in the plasma 
created in these collisions, 
(ii) provide a constraint for distinguishing between the two leading 
eccentricity models, {\em i.e.} the Glauber  \cite{Miller:2007ri} and the
factorized Kharzeev-Levin-Nardi (KLN) \cite{Kharzeev:2000ph,Lappi:2006xc,Drescher:2007ax} models, 
and (iii) point to an independent and robust method to estimate $\eta/s$.  

	The double differential data, $v^*_{n}(p_T,\text{cent})$ and $v_{n}(p_T,\text{cent})$, employed 
in our analysis are obtained from measurements by the PHENIX collaboration, for Au+Au collisions 
at $\sqrt{s_{NN}}$ = 0.2 TeV \cite{Adare:2010ux,Adare:2011tg}, and measurements by the ATLAS 
collaboration for Pb+Pb collisions at $\sqrt{s_{NN}}$ = 2.76 TeV \cite{JJia:2011hf,Atlas_Flow_Note}. 
The Au+Au data set exploits the event plane analysis method (c.f. Eq. \ref{eq:3}),
while the Pb+Pb data set utilizes the two-particle $\Delta\phi$ correlation technique
(c.f. Eq. \ref{eq:2}), as well as the event plane method. Note as well that, due to partial 
error cancellation, the relative systematic errors for the ratios $v_n/(v_2)^{n/2}$ 
and $v^*_n/(v_2)^{n/2}$ can be smaller than the ones reported for the $v_n$ values. 

%
%
\begin{figure}[t]
\includegraphics[width=1.0\linewidth]{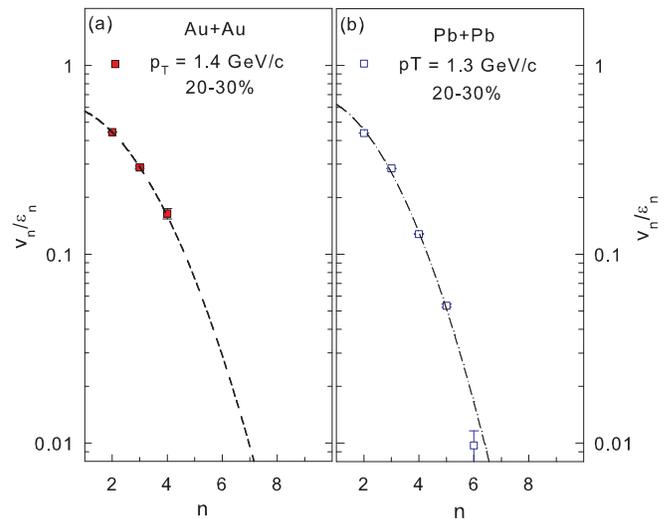}
\caption{$v_{n}/\varepsilon_n$ vs. $n$ for charged hadrons ($p_T \sim 1.4$ GeV/c) produced 
in Au+Au collisions at $\sqrt{s_{NN}}= 0.2$ TeV (a) and 
Pb+Pb collisions at $\sqrt{s_{NN}}= 2.76$ TeV. The $v_n$ data are taken from 
Refs. \cite{Adare:2011tg} and \cite{JJia:2011hf,Atlas_Flow_Note} respectively
for 20-30\% centrality.
The curves represent fits to the data (see text). 
}
\label{Fig1}
\end{figure}

	To perform validation tests for viscous damping compatible with sound 
propagation, the measured values of $v_n({\text{cent}})$, for each  
$p_T$ selection, were first divided by $\varepsilon_n({\text{cent}})$ and 
then plotted as a function of $n$. Monte Carlo (MC) simulations were used to 
compute $\varepsilon_n({\text{cent}})$ from the two-dimensional 
profile of the density of sources in the transverse  plane $\rho_s(\mathbf{r_{\perp}})$, 
with weight $\omega(\mathbf{r_{\perp}}) = \mathbf{r_{\perp}}^n$ \cite{Lacey:2010hw}. 
Figs.~\ref{Fig1} (a) and (b) show representative examples of $v_n/\varepsilon_n$ vs. $n$ for 
charged hadrons ($p_T \sim 1.4$~GeV/c) in mid-central Au+Au and Pb+Pb collisions
respectively. They confirm the exponential decrease of $v_n/\varepsilon_n$ 
with $n^2$, expected for sound propagation. This ``acoustic scaling'' of $v_n$ 
is further confirmed by the dashed and dot-dashed curves which indicate 
exponential/Gaussian fits $(Ae^{-\beta n^2})$ to the data shown. 

Similar patterns were observed for a broad selection of centralities for $p_T \alt 3$~GeV/c. 
However, for the 0-5\% 
and 5-10\% most central Pb+Pb collisions, $v_2/\varepsilon_2$ shows 
significant suppression relative to the empirical trend for $v_n$ vs. $n$, for other 
centralities shown by the curves in Fig.~\ref{Fig1}. The fractional magnitude of this suppression is essentially 
independent of $p_T$ even though $v_2/\varepsilon_2 < v_3/\varepsilon_3$ for $p_T \agt 2$ GeV/c.
We interpret this suppression as an indication that, for the most central Pb+Pb collisions, 
the low frequency modes $n < 3$ exceed the superhorizon limit, {\em i.e. $2\pi R_{\!f}/n > 2r_s$}.
That is, for these low frequency modes, the requirement for 
the maximum momentum anisotropy to develop is not met, so only a fraction of the full  
anisotropy is observed. For mid-central collisions, these sound modes have  
shorter wavelengths which make them sub-horizon. Note that $R_{\!f} = \bar{R} + r_s$, so 
the order $n$ of the low frequency modes which are suppressed, can serve to constrain 
the sound speed.

\begin{figure}[t]
\includegraphics[width=1.0\linewidth]{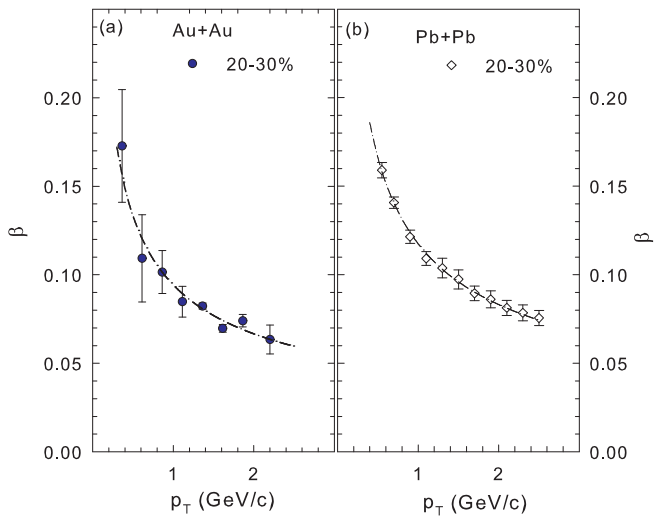}
%
\caption{$\beta$ vs. $p_T$ for 20-30\% central Au+Au collisions at $\sqrt{s_{NN}}= 0.2$ TeV (a) 
and 20-30\% central Pb+Pb collisions at $\sqrt{s_{NN}}= 2.76$ TeV (b). The dot-dashed and 
dashed curves indicate a $1/\sqrt(p_T)$ dependence for $\beta$ (see text).
}
\label{Fig2}
\end{figure}

	 For each centrality, exponential fits $( Ae^{-\beta n^2} )$ to $v_n/\varepsilon_n$ vs. $n$
were also made to investigate the $p_T$-dependent viscous corrections attributable to $\delta f(p_T)$.
Figs.~\ref{Fig2}~(a) and (b) show the $p_T$-dependence of the $\beta$ values 
extracted for 20-30\% central Au+Au and Pb+Pb collisions respectively; similar data trends 
were observed for other centralities. The dashed and dot-dashed curves in Fig. \ref{Fig2} show 
that $\beta$ scales as $1/\sqrt(p_T)$ for both collision energies, but the values for 
Pb+Pb collisions are about 25\% larger. This scaling is a clear  
indication of the influence of the relaxation time \cite{Dusling:2009df,Lacey:2010fe}. 
Consequently it should serve as an important constraint for models. 
%

\begin{figure}[t]
\includegraphics[width=1.0\linewidth]{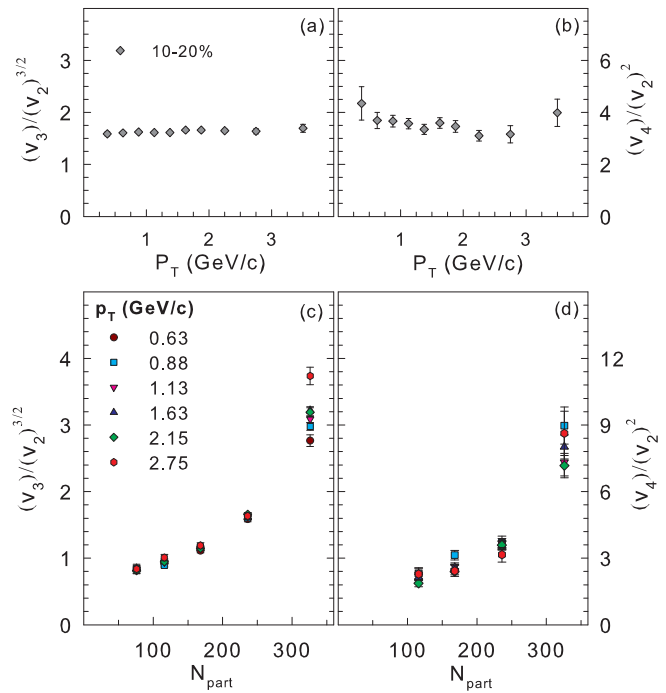}
\caption{$v_{3}/(v_2)^{3/2}$ vs. $p_T$ (a) and $v_{4}/(v_2)^{2}$ vs. $p_T$
for 10-20\% central Au+Au collisions. The bottom panels show 
$v_{3}/(v_2)^{3/2}$ vs. $N_{\text{part}}$ (c) and $v_{4}/(v_2)^{2}$ vs. $N_{\text{part}}$ (d)
for several $p_T$ cuts, as indicated. The $v_{2,3,4}$ values used for these ratios 
are reported in Ref.  \cite{Adare:2011tg}.
}
\label{Fig3}
\end{figure}

Figure \ref{Fig3} shows the ratios $v_3/(v_2)^{3/2}$ and $v_4/(v_2)^2$
plotted as a function of $p_T$ [(a) and (b)] and $N_{\text{part}}$ [(c) and (d)]
respectively, for Au+Au collisions. These ratios indicate an essentially flat dependence 
on $p_T$, but show a characteristic increase with $N_{\text{part}}$. 
The same trends are exhibited by the Pb+Pb data ($\sqrt{s_{NN}}$ = 2.76 TeV) with magnitudes 
comparable to those for Au+Au collisions for the same range of $p_T$ and centrality selections. 
We interpret the flat $p_T$ dependence of $v_{n}/(v_2)^{n/2}$ [for each ${\text{cent}}$] to be 
an indication that the $p_T$-dependent viscous corrections for $v_n$ are dominated by the 
$p_T$-independent contributions which stem from the dispersion relation for sound propagation. 

The trends for $v^*_4/(v_2)^2$ were found to be similar to those for $v_4/(v_2)^2$, but the 
ratios $v^*_4/(v_2)^2$ vs. $N_{\text{part}}$ are much less steep \cite{Adare:2010ux}.
The $N_{\text{part}}$ dependence of $v^*_4/(v_2)^2$ and $v_4/(v_2)^2$
contrasts with the constant value of $\sim 0.5$ predicted for perfect fluid 
hydrodynamics \cite{Borghini:2005kd,Csanad:2003qa}, and points to the important 
role of the higher-order eccentricity moments and their 
fluctuations \cite{Lacey:2009xx,Lacey:2010yg,Lacey:2010fe,Alver:2010gr,Lacey:2010hw}. 
The apparent differences between $v^*_4/(v_2)^2$ and $v_4/(v_2)^2$ are also an indication of  
the important role of $\varepsilon_4$ as a driver for $v_4$. That is, the expected contribution 
to $v_4$ from $v_2$ [$\sim (v_2)^2$] does not dominate the $v_4$ measurements. 
Note as well that $v_4 > v^*_4$ is expected because the initial eccentricity 
fluctuations cause $\Psi_2$ to fluctuate about $\Psi_4$.  

%

\begin{figure}[t]
\includegraphics[width=1.0\linewidth]{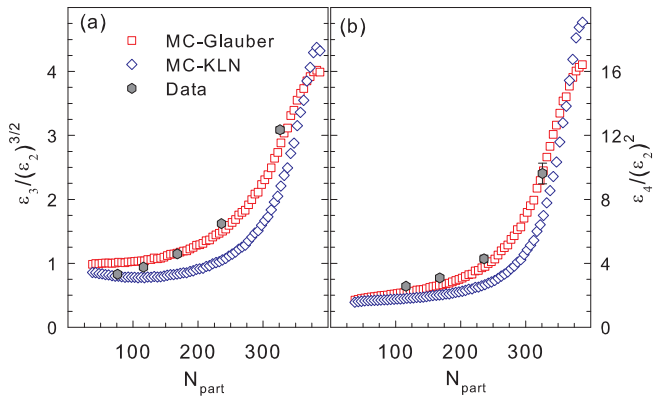} 
\caption{Data comparisons to the calculated ratios (a) $\varepsilon_3/(\varepsilon_2)^{3/2}$ vs. $N_{\text{part}}$ 
and  (b) $\varepsilon_4/(\varepsilon_2)^{2}$ vs. $N_{\text{part}}$ for MC-Glauber 
and MC-KLN initial geometries for Au+Au collisions (see text).
}
\label{Fig4}
\end{figure}

The flat $p_T$ dependence for $v_{n}/(v_2)^{n/2}$ (c.f Fig. \ref{Fig3}) also suggests 
that the $p_T$-dependent contributions to the viscous corrections for the ratios 
$(v_n/\varepsilon_n)/(v_2/\varepsilon_2)^{n/2}$ essentially cancel, making them 
a reliable constraint for the ratios $\varepsilon_n/(\varepsilon_2)^{n/2}$ and consequently, 
an important route for distinguishing between different eccentricity models \cite{Lacey:2010hw}. 
The solid symbols in Fig. \ref{Fig4} show a representative set of the experimental 
$v_n/(v_2)^{n/2}$ ratios which take account of the relatively small effects of acoustic 
suppression. The open symbols show the corresponding eccentricity ratios obtained for the two 
eccentricity models. The ${\varepsilon_{n}}$ values for these ratios were evaluated as described earlier. 
Fig. \ref{Fig4} indicates relatively good agreement between data and 
the $\varepsilon_n/(\varepsilon_2)^{n/2}$ ratios, 
confirming the utility of $v_n/(v_2)^{n/2}$ as a constraint for distinguishing between 
the eccentricity models \cite{Lacey:2010hw}.
%

	The observed scaling patterns summarized in Figs. \ref{Fig1} - \ref{Fig3} undoubtedly 
provide an important set of constraints for detailed comparisons to model calculations. 
In lieu of such calculations, we demonstrate their current utility for first rough 
estimates of the magnitude of $\eta/s$ and the viscous horizon. 
To this end, we employ fits to both the Au+Au and Pb+Pb data, with the 
fit function $(A\sqrt(p_T) e^{-\frac{\beta n^2}{\sqrt(p_T)}}$, 
where the $\sqrt(p_T)$ factors account for the influence of $\delta f(p_T)$.
These fits indicate that, relative to $v_2/\varepsilon_2$, the magnitude of 
$v_n/\varepsilon_{n, n\ge 3}$ decreases by more than a factor of $50$ 
for $n_v \sim 7$, {\em i.e.} $v_n/\varepsilon_n$ for $n \agt 7$ are essentially completely 
damped. This gives the estimate $r_v = 2\pi\bar{R}/n_v \simeq 1.8$ fm  for the 
viscous horizon in central Au+Au and Pb+Pb collisions.  

	The same fits allow robust extraction of the $\beta$ values for Au+Au and 
Pb+Pb collisions. Here, it is noteworthy that the dependence of $v_n/\varepsilon_n$ on 
$n$ can provide a particularly tight constraint, because it is the relative magnitudes 
of $v_n/\varepsilon_n$ which now serve to constrain $\beta$. These $\beta$ values 
have been used to extract a first rough estimate of $4\pi\eta/s \sim 1.2$ for central 
Au+Au collisions for $\left< T \right> = 220$ MeV and $t=9$ fm \cite{Dusling:2007gi}. 
This rough estimate is in reasonable qualitative agreement with the 
values from prior extractions~\cite{Lacey:2006bc,Adare:2006nq,Romatschke:2007mq,Luzum:2008cw,
Xu:2007jv,Drescher:2007cd,Song:2008hj,Lacey:2009xx,Dusling:2009df,Denicol:2010tr}. 
A similarly rough estimate from the Pb+Pb data ($\sqrt{s_{NN}}= 2.76$ TeV) gives a 
value for $\eta/s$ which is approximately 25\% larger (cf. the larger value for $\beta$) 
if  we assume that the ratio $T/t$ is roughly the same for Au+Au and Pb+Pb 
collisions \cite{Schenke:2011tv}. That is, we assume that a possibly larger flow 
freeze-out time is compensated for, by a higher estimated mean temperature. 
More detailed model calculations are required 
to address the values of $\left< T \right>$ and $t$ required for 
a more accurate estimate of $\eta/s$. Nonetheless, our procedure clearly demonstrates 
the value of the relative magnitudes of $v_n$ for extraction of $\eta/s$.


In summary, we have presented a detailed study of the scaling properties of the flow 
coefficients $v_n$ and their ratios $[v_n/(v_2)^{n/2}]_{n\geq 3}$, for 
Au+Au and Pb+Pb collisions at $\sqrt{s_{NN}}=0.2 \text{ and } 2.76$ TeV respectively. 
Within an empirically parametrized viscous hydrodynamical framework,
these properties can be understood to be a consequence of the acoustic nature of 
anisotropic flow, {\em i.e}, the observed viscous damping is characteristic of  
sound propagation in the plasma produced in these collisions. 
This interpretation not only provides a straightforward constraint for distinguishing 
between the two leading eccentricity models, it provides a means to constrain the 
relaxation time and the sound speed, as well as   
to make independent estimates for the the averaged specific shear viscosity and 
the viscous horizon, via studies of the relative magnitudes of $v_n$.  
The observed scaling also has important implications for accurate decomposition 
of flow and jet contributions to two-particle $\Delta\phi$ correlation 
functions. This is because the higher-order harmonics can be expressed as a 
power of the high precision $v_2$ harmonic. 
It will be valuable to perform detailed viscous hydrodynamical model comparisons 
to $v_n$ and $v_n/(v_2)^{n/2}$ for both identified and unidentified hadrons, as well 
as to establish the $p_T$ value which signals a breakdown of these  scaling
patterns.

{\bf Acknowledgments}
This research is supported by the US DOE under contract DE-FG02-87ER40331.A008. 
and by the NSF under award number PHY-1019387.
 


%
\bibliography{Ho_vn_pT_scaling} 
\end{document}